%% file: main.tex
\documentclass[sigconf]{acmart}

\AtBeginDocument{%
  }

\usepackage{graphicx} 

\usepackage{hyperref}
\usepackage{url}

\usepackage{amsfonts}       
\usepackage{nicefrac}       
\usepackage{microtype}      
\usepackage{xcolor}         
\usepackage{graphicx}
\usepackage{booktabs}
\usepackage{amsmath}
\usepackage[capitalize,noabbrev]{cleveref}
\usepackage{subcaption}
\usepackage{enumitem}
\usepackage{xspace}
\usepackage{mathtools}
\usepackage{amsthm}

\usepackage{thm-restate}

\newtheorem{theorem}{Theorem}[section]

\newtheorem{definition}[theorem]{Definition}

\usepackage[textsize=tiny]{todonotes}

\usepackage{algorithm}
\usepackage[noend]{algpseudocode}
\usepackage{wrapfig}

\usepackage{verbatim}

\usepackage{graphicx}
\usepackage{subcaption}
\usepackage{enumitem}

\usepackage{multirow}
\usepackage{placeins} 

\usepackage{afterpage}

\usepackage{float}

\DeclareMathOperator*{\argmin}{arg\,min}

\newcommand{\name}[0]{{\textsc{IP-DiskANN}}\xspace}
\newcommand{\longname}[0]{{\textsc{InPlaceUpdate-DiskANN}}\xspace}

\renewcommand\footnotetextcopyrightpermission[1]{}

\begin{document}

\title{In-Place Updates of a Graph Index for Streaming Approximate Nearest Neighbor Search}


\author{Haike Xu}
\email{haikexu@mit.edu}
\affiliation{%
  \institution{MIT}
  \country{}
}

\author{Magdalen Dobson Manohar}
\email{mmanohar@microsoft.com}
\affiliation{%
  \institution{Microsoft Azure and Carnegie Mellon University}
  \country{}
}

\author{Philip A. Bernstein}
\email{philbe@microsoft.com}
\affiliation{%
  \institution{Microsoft Research}
  \country{}
}

\author{Badrish Chandramouli}
\email{badrishc@microsoft.com}
\affiliation{%
  \institution{Microsoft Research}
  \country{}
}

\author{Richard Wen}
\email{rwen1@umd.edu}
\affiliation{%
  \institution{University of Maryland}
  \country{}
}

\author{Harsha Vardhan Simhadri}
\email{harshasi@microsoft.com}
\affiliation{%
  \institution{Microsoft Azure}
  \country{}
}


\begin{abstract}
Indices for approximate nearest neighbor search (ANNS) are a basic component for information retrieval
and widely used in database, search, recommendation and RAG systems.
In these scenarios, documents or other objects are inserted into and deleted from the 
working set at a high rate, requiring a stream of updates to the vector index. 
Algorithms based on proximity graph indices are the most efficient indices for ANNS,
winning many benchmark competitions. 
However, it is challenging to update such graph index at a high rate,
while supporting stable recall after many updates. 
Since the graph is singly-linked, deletions are hard because there is no fast way to find in-neighbors of a deleted vertex.
Therefore, to update the graph, state-of-the-art algorithms such as FreshDiskANN accumulate deletions
in a batch and periodically consolidate, removing edges to deleted vertices and modifying the graph
to ensure recall stability. 
In this paper, we present \name (\longname), the first algorithm to avoid batch consolidation by efficiently
processing each insertion and deletion in-place.
Our experiments using standard benchmarks show that \name has stable recall
over various lengthy update patterns in both high-recall and low-recall regimes.
Further, it's query throughput and update speed are better
than using the batch consolidation algorithm and HNSW. 
\end{abstract}





\maketitle
\pagestyle{plain}

\input{intro}

\input{prelims}
\input{algos}
\input{experiments}
\input{ablation}
\input{related}

\input{conclusion}

\bibliographystyle{ACM-Reference-Format}
\bibliography{ref}

\clearpage
\newpage



\end{document}

%% file: intro.tex
\section{Introduction}

Today, it has become commonplace to represent text,  web pages, images,
audio,  video and multi-modal documents as \emph{vectors} in high-dimensional vector spaces using deep learning models. 
These models represent various notions of semantic similarity as distances between
vector representations of the objects, so that the nearest points to a query are the
most semantically relevant~\cite{clip, ance, trecDL22}.
An \emph{approximate nearest neighbor search} (\emph{ANNS}) a.k.a. \emph{vector search}
system indexes and queries such vectors on collections of scales ranging from thousands
to hundreds of billions of vectors.
ANNS systems are used in online real-time applications and services such as search,
recommendation, and retrieval-augmented generation (RAG)~\cite{rag}. 
The latter is a core mechanism to augment a large language model (LLM) with content that
the LLM was not trained on, such as an individual's or organization's private content~\cite{copilot}.
Given a user query, the ANNS helps find relevant private content to add to the input prompt
to improve the quality of the LLM's response.
RAG systems are also used to ground LLMs and aid their reasoning capabilities with factual content
(e.g., OpenAI DeepResearch~\cite{deepresearch}).

Data-agnostic indexing methods based on Locality-Sensitive Hashing (LSH)~\cite{Indyk98, LSHSurvey08, Broder98}
have theoretically optimal trade-offs between accuracy (ratio of distance to retrieved points
vs. optimal result) and query complexity (number of points in the index that the query is compared to).
However, data-dependent indices based on clustering~\cite{IMI12, douze2024faiss, SPANN}
and proximity graphs~\cite{HNSW16, RNG, iwasaki2018, DiskANN19} are more
efficient than the best implementations of LSH such as ANNOY~\cite{github:annoy}.
Proximity graphs in particular have been found to be very efficient in practice, with  their implementations
providing a good trade-off across accuracy, query latency, and  throughput~\cite{Benchmark, bigann21, bigann23}.
They are also adaptable to variations of ANNS that arise in practice, such as sparse high-dimensional indices
and filtered queries ~\cite{bigann23, FilteredDiskANN, acorn}.

A proximity graph consists of a designated starting point in the high-dimensional space,
nodes representing each vector, and directed edges that connect these nodes. 
The edge set of a proximity graph is carefully designed such that the closest neighbors of 
a query can be found via a greedy search of the graph which, at each node, walks towards
the out-neighbors closest to the query (see Algorithm~\ref{alg:search}).
The edges consist a mix of nearby as well as some far-away vectors strategically selected to accelerate search.
There are various techniques to build such graphs with varying graph construction cost,
search quality, and efficiency. 
In this paper, we focus on the simple yet powerful graph index in the DiskANN system~\cite{DiskANN19, diskann-github}.
This technique is used in entries in recent NeurIPS BigANN competitions~\cite{bigann21, bigann23},
and is reported to be used extensively in global services such as Bing and Microsoft 365,
edge devices via Windows Copilot runtime~\cite{windowscopilot},
and databases~\cite{cosmosdb, pgvectorscale, JVector}.

While proximity graphs have been studied in the \emph{static} setting,
there is a strong need for \emph{streaming updates} of vector indexes. 
Online services continuously generate new contents from sources such as
 database records, emails, videos, tweets, photographs, and webpages. 
They also \emph{update} existing content by editing 
existing documents, or deleting content due to expiration, inaccuracy, or privacy reasons. 
Such updates should be made be available to new queries immediately,
so that the ANNS system serves fresh content at all times. 
This means that vector indices cannot be treated as static data,
and must be updated while maintaining high quality in terms of query recall and latency.

While insertion of new vectors is not too hard 
(since many build algorithms for proximity graphs
are naturally incremental),
maintaining index quality after deletions 
of vectors is quite challenging.
Replacements of vectors are tricky to handle as well.
As a result, many vector indices resort to \emph{soft-deletion} or \emph{segment-based} strategies
both of which require expensive periodic index rebuilds.

In the soft-deletion based strategy, indices mark each deleted point with a tombstone but do not
remove it from the graph. As a result query quality degrades over time, and indices
are typically rebuilt after, say, 10-20\% of nodes are deleted. 

In the segment-based strategy, updates are captured in \emph{update} files which are published
as immutable segments after they reach a certain size. 
A vector index is built for the content in the index~\cite{Lucene-vector}.
Periodically, segments are merged and indices are rebuilt.
A query is processed using a combination of segment-level indices and a full scan of the update file which can be inefficient. Lucene~\cite{Lucene} has recent GitHub issues created on the CPU~\cite{lucene-hnsw-cpu} and memory~\cite{lucene-hnsw-memory} overheads of HNSW~\cite{HNSW16} graph merges, which indicate this to be a problem in practice.


With either strategy, this leads to expensive rebuilds that can be impractical for large indices.
For example, building a DiskANN index for one  billion scale SIFT vectors incurred a peak
memory usage of 1.1TB and took 2 days with 32 vCPUs~\cite{DiskANN19}.

An improvement on this is an incremental rebuild approach, as in FreshDiskANN~\cite{freshdiskann},
the state-of-the-art update algorithm for DiskANN.
Instead of rebuilding the index from scratch, it relies on incremental merges of tiered segments.
A novel algorithm merges  recent updates and smaller lower-tier segments into larger higher-tier segments
to produce a new index with minimal write amplification.
Although this is more efficient than rebuilding the index from scratch,
it still suffers from the other problems.
For example, usage spikes from such rebuilds increase the 99.9$^\textrm{th}$ percentile of query latency.
Moreover, the system must be provisioned with resources to handle usage spikes during merge,
which increases the overall cost of running the ANNS system. 
Further, queries are less efficient due to the need to query indices over multiple segments.

What we want is a way to process updates to the ANNS graph index \emph{in place}.
The update algorithm must be fast and produce a graph that supports high recall.
It must also be stable over long periods, so there is no need to rebuild the index.
This would enable the ANNS system to maintain a single graph and absorb updates
in a streaming fashion. It would avoid the periodic spikes in resource usage and
query latency due to rebuild, which would translate to a stable and low-cost
online service offering in the cloud.

In-place insertions can use the same algorithm as the one used to build the graph.
Deletions are more problematic.
Since the index graph is singly-linked with out edges,
when deleting a vertex we usually do not know its in-neighbors. 
An obvious fix is to doubly-link the graph. 
However, since most of the space consumed by a vertex is for its list of neighbors, 
this would either halve the number of vectors that can be handled by a single graph or double its cost.
It would also make the system more unwieldy due to increased locking complexity
in a highly concurrent system. In practice, all these reasons make maintaining
in-neighbors unacceptable.


\textbf{Our main contribution is \name, a new algorithm for in-place deletions in a DiskANN graph index},
thereby avoiding periodic batch consolidation or rebuilds. Furthermore, \name has the
following properties:
\begin{itemize}
\item Queries over are as efficient as those over a graph built from scratch using the active vectors.
\item Maintains stable recall over long update sequences of various kinds.
\item The cost of each update in the stream
is better than the cost of vector insertion during a static index build,
and compared to prior lazy approaches.
\item The cost of updates is also significantly lower than for 
the HNSW algorithm, while providing comparable or better recall.
\end{itemize}

The paper is organized as follows:
Section~\ref{sec:prelim} formally defines the notation related metrics used in this paper.
Section~\ref{sec:algorithms} presents our algorithm, \name.
An experimental evaluation is in Section~\ref{sec:experiments} followed by an ablation study in Section~\ref{sec:ablation}.
Section~\ref{sec:related} discusses related work, and Section~\ref{sec:conclusion} concludes the paper.



%% file: prelims.tex
\section{Preliminaries and Notation}
\label{sec:prelim}
We define the streaming nearest neighbor problem as follows.
We are given a time-varying dataset of \emph{points}, where $P_t$ denotes the dataset at time $t$.
Each point $p$ has an associated vector, denoted by subscripted variable, such as $x_p$.
We sometimes use the terms point and vertex interchangeably, since they are in 1:1 correspondence.

The algorithm needs to support three operations:
\begin{itemize}
\item Insert a point $p$ to the current point set $P_t$.
\item Delete an existing point $p$ from the current point set $P_t$.
\item Search for the nearest neighbors of a given query $q$ in the current point set $P_t$.
\end{itemize}

The degree of the graph, i.e., maximum number of out-neighbors of a node in the graph, is denoted $R$.
We measure the accuracy of the index over $P_t$ for a query using the following notion of recall.
\begin{definition}[Recall@k]
Given a point set $P_t$ and query $q$, let $G\subseteq P_t$ be the set of actual nearest $k$ points to $q$,
and $A\subseteq P_t$ be the set of $k$ vectors returned by an ANN algorithm.
The Recall@k for this algorithm in answering the query is defined as $\frac{|G\cap A|}{k}$.
\end{definition}
As is the standard practice, we report recall@10 in this paper.

The performance of an index is measured by:
\begin{itemize}
\item the recall-query complexity trade-off for query answering.
Query complexity can be measured in terms of the latency (or throughput)
of an implementation on certain machine. It can also be measured in a hardware
and implementation-agnostic way by counting the number of points in the
index that the query is compared with. We report whichever is best suited
to the experiment's context.
\item the complexity of handling insertions and deletions.
\end{itemize}

We denote a graph-based index over point set $P_t$ as the directed graph $G(P_t,E_t)$ with slight overloading of notation.
Each vertex in the graph corresponds to a vector in $P_t$. 
For each point $p\in P_t$, we use $N_{out}(p)$ to denote the out-neighbors of vertex $p$
and $N_{in}(p)$ to denote its in-neighbors.
The in-neighbors will not necessarily be maintained in the algorithms, but are useful in the analysis.

We use $x_q$ to denote the co-ordinates of a query vector $q$.
We use $d(x_p,x_q)$ to denote the distance between vectors $p$ and $q$.
For a set of vertices $V$, we use $|V|$ to denote the size of the vertex set.

%% file: algos.tex
\section{Algorithms}
\label{sec:algorithms}

Before introducing our in-place deletion algorithm,
we recap basic ideas from the DiskANN and FreshDiskANN algorithms which we use as our baseline.
We describe ideas behind searching, incremental indexing, and consolidation of changes to the index.

\subsection{DiskANN}

It is easier to describe the  construction of the DiskANN index once we understand how
how to answers queries, since the index must be built to support queries efficiently.
In fact, the following Greedy Search query algorithm is a subroutine in index construction.

\textsf{{\bf GreedySearch} (Algorithm~\ref{alg:search})} 
To find the nearest neighbors of the query vector $q$, the algorithm performs 
a greedy beam search  on the graph starting from a fixed starting point $s$.
At each step, the algorithm \emph{expands} the closest unexplored vertex,
computes the distance from $x_q$ to each out-neighbor, and adds the neighbors
to a priority queue sorting on increasing distance from $x_q$.
The search iteratively does this operation until all the top-$l_s$
closest vertices in the priority queue have been explored. 
At the end, the algorithm returns the top-k closest vertices seen as the answers.

\begin{algorithm}[h]
\caption{GreedySearch($G$, $x_q$, $k$, $l_s$)}
\label{alg:search}
\begin{algorithmic}[1]
\State \textbf{Input}: Current index graph $G(P,E)$, query $x_q$, search parameter $l_s$.
\State \textbf{Output}: Visited list $V$ and top-$k$ nearest neighbors

\State Initialize $L=\{s\}$ and $V=\{\}$ // $s$ is the starting point of $G$.

\While{$L\setminus V\neq \varnothing$}
\State $v\gets\argmin\limits_{v\in L\setminus V} D(x_v,q)$
\State $L\gets L\cup N_{out}(v)$
\State $V\gets V\cup v$
\If{$|L|>l_s$}
\State $L\gets$ top $l_s$ closest vertices to $q$ in $L$
\EndIf
\EndWhile
\State \textbf{Return} $V$ and $k$ closest vertices to $q$ in $V$
\end{algorithmic}
\end{algorithm}

\textsf{{\bf Insert} (Algorithm~\ref{alg:insert}):}
The DiskANN graph is built incrementally by inserting the vectors into the graph.
\footnote{Although the original paper~\cite{DiskANN19} describes a 2-pass construction,
in practice one pass suffices~\cite{diskann-github} and also lowers index construction
time normalized for index quality.}
To insert a new vector $x_p$, it searches for $x_p$ using \textsf{GreedySearch} 
and captures the list of expanded vertices, $V$.
It then \emph{prunes} $V$ down to a list $V'$ of cardinality $\leq R$,
and attempts connections between every $v\in V'$ and $p$ in both directions. 
If this causes the degree of vertex $v$ to exceed $R$, it prunes $N_{out}(v)$
to reduce the out-neighbor list to below the degree limit. 

\begin{algorithm}
\caption{Insert($G$, $x_p$,  $l_b$, $R$, $\alpha$)}
\label{alg:insert}
\begin{algorithmic}[1]
\State \textbf{Input}: Current index graph $G(P,E)$, point to be inserted $x_p$, build parameter $l_b$, graph degree $R$, pruning parameter $\alpha$
\State \textbf{Output}: Graph $G(P',E')$ where $P'=P\cup \{p\}$

\State $[V,A]\gets$ \textsf{GreedySearch}($G$, $x_p$, $1$, $l_b$)
\State $N_{out}(p)\gets$ \textsf{RobustPrune}($G$, $p$, $V$, $R$, $\alpha$)
\For{$v\in N_{out}(p)$}
\State $N_{out}(v)\gets N_{out}(v)\cup {p}$
\If{$|N_{out}(v)|>R$}
\State $N_{out}(v)\gets$ \textsf{RobustPrune}($G$, $v$, $N_{out}(v)$, $R$, $\alpha$)
\EndIf
\EndFor

\end{algorithmic}
\end{algorithm}

\textsf{{\bf RobustPrune} (Algorithm~\ref{alg:pruning}): }
When a vertex $p$ exceeds the degree limit $R$, \textsf{RobustPrune} selects the edges to retain. 
Let $V$ be the  list of $p$'s out-neighbors, sorted by their distance to vector $x_p$. 
At each step, remove the point $v\in V$ that is currently closest to $x_p$,
add $v$ to the new neighbor list $V'$, and remove those points still in $V$  that
are much closer to $v$ than to $p$ from $V$. 
The algorithm repeatedly adds edges to the new neighbor list until it reaches the degree limit.
The prune procedure determines the structure of the graph.
The parameter $\alpha$
is carefully set so that prune retains a mixture of nearby nodes and directionally diverse distant nodes.
It is set to constant slightly greater than $1$, typically $1.2$,
to reduce the number of hops seen by greedy search~\cite{DiskANN19}.

\begin{algorithm}
\caption{RobustPrune($G$, $p$, $V$, $R$, $\alpha$)}
\label{alg:pruning}
\begin{algorithmic}[1]
\State \textbf{Input}: Current index graph $G(P,E)$, point $p$, 
candidate out neighbors $V$, graph degree $R$, pruning parameter $\alpha$.
\State \textbf{Output}: A subset $V'\subset V$ of cardinality $\leq R$.
\State $V'=\varnothing$
\While{$V\neq \varnothing$}
\State $v\gets \argmin\limits_{v\in V}D(x_v,x_p)$
\State $V'\gets V'\cup v$
\State $V\gets V \setminus v$
\State $V\gets \{u\in V: \alpha\cdot D(x_u,x_v) > D(x_u,x_p)\}$
\EndWhile

\end{algorithmic}
\end{algorithm}

\subsection{FreshDiskANN}

While DiskANN naturally supports insertion of new vertices, there is no trivial way to delete vectors in place.
The \emph{lazy} deletion method introduced by FreshDiskANN \cite{freshdiskann} marks deleted
vertices online. The \textsf{GreedySearch} procedure still navigates vertices marked as deleted,
but does not return them as part of the visited list, or as nearest neighbors in the final answer. 
Periodically, the graph is \emph{consolidated} in the background to clean up the deleted vertices,
while preserving the index quality, as follows:

\textsf{{\bf Consolidation} (Algorithm~\ref{alg:consolidation_baseline}):}
Let $D$ be the set of deleted vertices. The background consolidation eliminates the vertices $D$
from the graph. It also cleans up the  neighborhood of each $p\in D$ to preserve the graph's \emph{navigability}
by (a) adding an edge from each up-stream neighbor of $p$ (i.e., $N_{in}(p)$) to each downstream neighbor (in $N_{out}(p)$),
and (b) RobustPruning each $v\in N_{in}(p)$ that has exceeded their degree limit.
Since  $N_{in}(p)$ is not explicitly maintained, the same effect can be had by
iterating through each $p\in P$, finding $p$'s deleted out-neighbors
$C = \{v\in N_{out}(p) \cap D\}$, and concatenating their out-neighbors  to $N_{out}(p)$, i.e.,
$N_{out}(p) \gets N_{out}(p) \cup_{v\in C} N_{out}(v) \setminus D$.
Pruning is done on $N_{out}(v)$ if degree exceeds bounds,
and $\alpha>1$ (e.g, $1.2$) is critical for maintaining recall.

\begin{algorithm}
\caption{Consolidation($G$, $D$, $R$, $\alpha$) (baseline)}
\label{alg:consolidation_baseline}
\begin{algorithmic}[1]
\State \textbf{Input}: Current index graph $G(P,E)$, deleted point set $D$.
\State \textbf{Output}: Updated index graph $G$ on nodes $P'=P\setminus D$.

\For{$p$ in \textsf{$P \setminus D$}}
    \State $C\gets N_{out}(p)\cap D$
    \For{$v$ in $C$}
        \State $N_{out}(p)\gets N_{out}(p) \cup (N_{out}(v)\setminus D)$ 
    \EndFor
    \State $N_{out}(p)\gets$ RobustPrune($G$, $p$, $N_{out}(p)$, $R$, $\alpha$)
\EndFor

\end{algorithmic}
\end{algorithm}

\subsection{In-place deletion algorithm}

In this section, we present the main algorithmic ideas for handling deletions in place.
First, let us consider the motivation behind the method used in FreshDiskANN and its limitations.
To delete a vertex in a graph without affecting connectivity, a simple approach is to
connect all its in-neighbors to all its out-neighbors. However, this impacts graph quality in two ways:

\begin{itemize}
\item  A deleted vertex with $R$ in-neighbors and $R$ out-neighbors
will result in $R^2$ edges added to the graph. This can be slow since it 
could trigger too many invocations of the robust pruning procedure to select edges. 
\item Some unnecessary short range edges will be added forcing out a few useful long range ones
due to the degree limit, degrading the overall connectivity.
\end{itemize}

In Algorithm~\ref{alg:deletion}, we propose a solution that remedies these two issues.
Instead of connecting all vertices in $N_{in}(p)$ to all vertices in $N_{out}(p)$, we could augment the out-neighborhood of each $z\in N_{in}(p)$ with at most $c$ points near $x_p$ and $x_z$.
Similarly, for each $w\in N_{out}(p)$, we can find at most $c$ existing vertices near $x_p$ and $x_w$,
and add $w$ to their out-neighborhood.
The constant $c$ has to be small so that the number of added edges $O(cR)$ is much smaller than $O(R^2)$.
Yet, too small a value would impact the navigability of the graph.
We found $c=3$ a reasonable trade-off (more in Sec.~\ref{sec:ablation}).

Another issue is that we cannot easily enumerate the in-neighbors of the deleted point
as they are not explicitly stored in practical implementations.
The FreshDiskANN algorithm bypasses this issue by adding connections only
during an offline consolidation phase, where it performs a scan over the index,
avoiding the need for in-neighbors. However, this approach requires
periodic heavy consolidations to maintain satisfactory graph quality,
which our in-place algorithm aims to avoid.

\begin{algorithm}[b]
\caption{In-place Deletion($G$, $p$, $l_{d}$, $k$, $c$, $\alpha$, $R$)}
\label{alg:deletion}
\begin{algorithmic}[1]
\State \textbf{Input}: Current index graph $G(P,E)$, the point to be deleted $p$,
deletion parameter $l_d=128$, candidate list size $k$, number of edge copies $c$, pruning parameter $\alpha > 1$, graph degree $R$.
Constants typically set to $k=50$,  $c=3$, $\alpha=1.2$.
\State \textbf{Output}: Updated index graph $G$ on nodes $P'=P\setminus \{p\}$.
\State [\textsf{Visited}, \textsf{Candidates}] $\gets$ GreedySearch($G$, $x_p$, $k$, $l_{d}$)
\State $N'_{in}(p)\gets \{z \in \textsf{Visited}: p \in N_{out}(z) \} $
\For{$z \in N'_{in}(p)$}
    \State $C_z \gets$ closest-$c$ points to $x_z$ in \textsf{Candidates}
    \State $N_{out}(z) \gets N_{out}(z) \cup C_z \setminus \{p\}$
\EndFor

\For{$w \in N_{out}(p)$}
    \State $C_w \gets$ closest-$c$ points to $x_w$ in \textsf{Candidates}
    \For{$y \in C_w$}
        \State $N_{out}(y) \gets N_{out}(y) \cup \{w\} $
    \EndFor
\EndFor

\State Remove vertex $p$ from $G$ (immediately)
\State For any updated vertex $v$ that exceeds degree bound $R$, $N_{out}(v)\gets$ \textsf{RobustPrune}($G, v, N_{out}(v), R, \alpha$)
\end{algorithmic}
\end{algorithm}

Instead, we find an approximation of $N_{in}(p)$ as follows.
During the insertion of point $x_p$, the visited list $V$ served as the source of in-neighbors of $p$.
Therefore, it is reasonable to expect that if we search for $x_p$ once more (when we aim to delete $p$),
the new visited list approximates the old one. 
Therefore, our in-place algorithm (Algorithm~\ref{alg:deletion}) runs \textsf{GreedySearch} to
obtain the visited list $V$. Those nodes in $V$ with an out-edge to $p$ are our approximation of $N_{in}(p)$.

We now consider how to find $c$ appropriate edges to add to $z\in N_{in}(p)$.
A good replacement for edge $(z,p)$ is one that is close to both endpoints. 
Therefore, among the \textsf{candidates}, which contains the closest $k$ points to $x_p$,
we select $C_z$, the top-$c$ closest points to $x_z$, and add edges from $z$ to $C_z$.
Similarly, for each $w \in N_{out}(p)$, we find a replacement for $(p,w)$ as follows:
search for the $C_w$, the closest $c$ nodes to $x_w$ in the visited set. 
Add edges from each member of $C_w$ to $x_w$.
These modifications end up adding more edges than those removed, and could
cause a few vertices to exceed the degree bound. We address this problem the same
way as in previous algorithms -- invoke \textsf{RobustPrune} on its out-neighbor list.
It is important to use a parameter $\alpha>1$ to preserve the navigability of the graph
(here we use $1.2$ as in prior work).

After deleting $p$ from $G$, there may still be dangling edges in $G$ pointing to $p$
since Algorithm~\ref{alg:deletion} does not guarantee all incoming edges to $p$ are deleted.
To address this problem, we periodically scan the index to remove such dangling edges using Algorithm~\ref{alg:consolidation_ours}
usually after the number of deletions exceed a certain fraction of the index size.
Although our algorithm for handling deletion is not perfectly online, this offline
consolidation procedure is extremely lightweight and does not do any distance calculations.
As a result,  it does not impose a significant burden on our underlying system.


\begin{algorithm}
\caption{Consolidation($G$, $D$) (ours)}
\label{alg:consolidation_ours}
\begin{algorithmic}[1]
\State \textbf{Input}: Current index graph $G(P,E)$, deleted point set $D$.
\State \textbf{Output}: Updated index graph $G$ on nodes $P'=P\setminus D$.

\For{$p$ in \textsf{$P \setminus D$}}
    \State $N_{out}(p)\gets N_{out}(p)\setminus D$ 
\EndFor

\end{algorithmic}
\end{algorithm}

%% file: experiments.tex
\section{Experiments}
\label{sec:experiments}
In this section, we demonstrate that our \name algorithm can preserve index quality
over a range of streaming scenarios that we emulate with the help of \emph{runbooks}.
A \emph{\textbf{Runbook}} is a collection of vectors and a sequence of updates---insertions, deletions, and 
replacements---that an index must perform using vectors from this collection. 
We measure the quality of the index by querying the index after every step in this sequence.
Runbooks can differ in:
\begin{itemize}[leftmargin=*]
\item The number of steps between inserting and deleting a vector.
\item Spatial correlation of vectors deleted in one step.
\end{itemize}

Taking these two factors into consideration, we designed the following three categories of runbooks:

\begin{itemize}[leftmargin=*]
    \item{SlidingWindow:} Randomly divide the dataset into $T_{max}=200$ parts of equal size.
    Insert one part per step from $T=1$ to $T_{max}$.
    Starting from $T=T_{max}/2+1$, delete points inserted $T_{max}/2$ steps ago.
    Each point is in the index for the same number of steps.
    The index size becomes stable starting from time $T_{max}/2+1$. 
    We measure the algorithm's query performance starting from step $T_{max}/2+1$.
    This runbook simulates  ``news'' indices that index recent data that expires after a certain time window.
    \item{ExpirationTime:} Points differ on how long they reside in the index:
    forever, for a long time, or for short period.  If the runbook is $T_{max}=100$ steps long,
    these could correspond to a lifespan of 100, 50, and 10 steps respectively. We set
    their proportion to 1:2:10, so we have roughly equal number
    of points of each type at any time.
    At each step, we insert a $1/T_{max}$ fraction of the dataset
    in random order, assign them a random class upon arrival,
    and delete previous points when they expire.
    This runbook tests how well an index can handle points with different lifetimes.
    \item{Clustered \cite{simhadri2024resultsbigannneurips23}:} We partition the dataset into 64 clusters using k-means clustering.
    The runbook consists of 5 rounds. In each round, we iterate all the clusters and insert
    a random proportion of the cluster to the index. Next, we iterate over all the clusters
    and delete a random proportion of the active points in the cluster from the index.
    In this runbook, nearby points are always inserted or deleted at the same time,
    and we measure the average query recall after handling each cluster. This runbook
    tests the index quality against extreme changes in the distribution of the active point set. This is the most challenging of the three types.
\end{itemize}

We can instantiate each type of runbook with a different dataset to cover
various embedding models and distance functions.
We instantiate with the two datasets below.
The runbook name ``MSTuring-10M-SlidingWindow''
indicates that it uses the Sliding Window template
with 10 million vectors from the MSTuring dataset.
We will release links to the runbooks' source after the anonymous review.
\begin{tabular}{c|c|c|c}
\toprule
  Dataset    & Dimensions  & Distance      & Embedding \\ \hline
  MSTuring   & 100         & Euclidean     & MS Turing~\cite{bigann21}  \\ \hline
  Wikipedia  & 768         & Inner Product & \href{https://cohere.com/embed}{Cohere} \\ \hline
\end{tabular}

\paragraph{Parameters}
We use the same parameters for the \name algorithm,
and the FreshDiskANN baseline we test it against.
Typically this is $R=64$, $l_b=l_s=128, \alpha=1.2$ for the high-recall regime.
and $R=32, l_b=l_s=64, \alpha=1.2$ for the low-recall regime. 
Consolidation is triggered for both algorithms when $20\%$ of the points in the graph are marked deleted.
For those parameters unique to our algorithm, we present ablation studies in Section~\ref{sec:ablation}.
In this section, we set them to $l_d=128$, $k=50$, $c=3$.

We also compare with the canonical implementation of
another popular graph-based ANN algorithm -- HNSW~\cite{HNSW-git}, using commit \texttt{c1b9b79}.
We use \textsf{M}=48, and \textsf{ef\_construction}=\textsf{ef\_search}=128.
These parameters are chosen so that they consume roughly the same space as our algorithm.

\paragraph{HNSW update logic}
We note that the above version of HNSW handles deletions by marking them deleted
and avoiding returning them for queries (see Github issues
\href{https://github.com/nmslib/hnswlib/issues/249}{249} and
\href{https://github.com/nmslib/hnswlib/issues/275}{275}).
Furthermore, incoming inserts may be used to ``replace" deleted
nodes to keep the size of the graph from overflowing some pre-set cutoff.
First, it updates all of the deleted point $p$'s one-hop neighbors
by adding all of $p$'s two-hop neighbors to each of them,
and then trimming them back down to respect the degree limit, first by retaining only the closest $ef\_construction$ to each node, then calling their prune process to retain at most $M$ neighbors.
Then, it proceeds like a standard insert, calculating new
edges for the replaced points and overwriting the old edges.
In our experiments, we called the replace operation whenever
20\% or more of the nodes in the graph were marked deleted,
consistent with how often we called the consolidate function in FreshDiskANN.

\paragraph{Machines and Code}
All our experiments were conducted using 16 threads on a 32 vCPU Azure Standard\_L32s\_v3 VM,
with an Intel® Xeon® Platinum 8370C (Ice Lake) processor with hyper-threading
and 256 GB DRAM. We used a closed-source Rust implementation that
is consistent with the open-source implementation of FreshDiskANN 
used in the 2023 NeurIPS Big ANN Benchmarks (v0.5.0.rc3.post1 of ~\cite{diskann-github}).
Our implementation is faster, includes the new \name algorithm,
and has an improved implementation of the inner product distance function.

\begin{table}[t]
\footnotesize
\begin{tabular}{p{1.5cm}|c|c|c|c|c}
\toprule
                                       &                 & Deletion & Insertion & Search & Recall@10 \\ 
                                       
\midrule
\multirow{3}{*}{\shortstack{MSTuring-10M \\ SlidingWindow}} 
                                    & FreshDiskANN         & 1710     & 1452      & 143    & 94.4   \\  
                                    & \name & 1110     & 1461      & 136    & 94.8   \\
                                    & HNSW & N/A & 4016     & 66    & 91.8   \\
                                      
\midrule
\multirow{3}{*}{\shortstack{MSTuring-30M \\ Clustered}} 
                                    & FreshDiskANN         & 3611     & 3187      & 845    & 91.8   \\
                                    & \name & 3146     & 3133      & 807    & 92.5   \\
                                    & HNSW & N/A & 15296     & 859    & 89.5   \\

\midrule
\multirow{3}{*}{\shortstack{Wiki-1M \\ ExpirationTime}}     
                                    & FreshDiskANN         & 38      & 88       & 63    & 94.8      \\  
                                    & \name & 92       & 100       & 71     & 97.1   \\ 
                                    & HNSW & N/A & 240     & 42    & 95.9   \\

\midrule
\multirow{2}{*}{\shortstack{Wiki-10M \\ ExpirationTime}}  & FreshDiskANN         & 3222      & 2633       & 148    & 97.9      \\  
                                       & \name & 2170       & 2592       & 141     & 97.9  \\ 

\midrule
\multirow{2}{*}{\shortstack{Wiki-35M \\ ExpirationTime}}  & FreshDiskANN         & 14535      & 12347       & 619    & 97.3      \\ 
                                       & \name & 10027       & 11886       & 584     & 97.5   \\ 
                                       
\bottomrule
\end{tabular}
\caption{Comparison of running time (in seconds) between \name,
FreshDiskANN and HNSW in the high-recall regime. 
We report recall@10, the total time spent
on deletion (including consolidation), insertion, and executing queries across different runbooks.
The insertion time for HNSW reflects all updates including deletions.
\vspace{-10pt}
}
\label{tab:running_time_highrecall}
\end{table}

\subsection{Observations}
We report the recall, number of distance computations per query, and query throughput
in the high-recall regime for five runbooks, at each step,
in Figure~\ref{fig:highrecallplots}. 
The \name algorithm consistently outperforms the FreshDiskANN baseline by an
average recall improvement of 0.4, 0.7, 2.3, 0.03 and 0.2 percentages over these runbooks,
while using fewer distance computations and providing higher query throughput.
This phenomenon is interesting as the same $l_s$ parameter is used in both algorithms
and one might expect queries to have the same complexity in both algorithms.
One possible explanation is that since \name adds $cR$ rather than $O(R^2)$
edges per deletion, we have fewer unnecessary edges. This results in a slightly sparser
graph which is faster to query.

We report the time each algorithm spends on handling insertion and deletion in Table~\ref{tab:running_time_highrecall}. We observe that though the average
insertion time and search time is roughly the same, \name spends less time
on handling deletions on large datasets
(e.g., for dataset with more than 10 million points) compared with FreshDiskANN and HNSW.
Here the deletion time for \name includes Algorithm~\ref{alg:deletion} 
as well as periodically cleaning up of dangling edges using Algorithm~\ref{alg:consolidation_ours}.
For FreshDiskANN, the deletion time is dominated by the background
consolidation phase. For HNSW, deletion time is included in insertion time, since deletion is implemented as a replacement of a deleted node with a newly inserted node. 

\include{highrecallplots}

We observe that HNSW's overall insertion and deletion cost tends to be significantly
higher than that of \name and FreshDiskANN. We hypothesize that this is due to
several mutually reinforcing factors.
First, since HNSW's deletion adds all the two-hop neighbors of the deleted point
to all its one-hop neighbors, it adds $O(R^3)$ edges per deletion,
compared with $O(cR)$ for \name and $O(R^2)$ for FreshDiskANN.
Second, since deletion can only be processed as replacement of
deleted points by newly inserted nodes, the steady size of the
index will be larger on average than that of FreshDiskANN and \name,
meaning that each operation takes somewhat longer.
This interacts especially poorly with the clustered runbook,
in which long runs of deletions are followed by long runs of insertions.
Finally, the lack of amortization in the consolidation process means that
non-deleted nodes might be pruned up to tens of times for a set of
deletions of 20\% of the index size, as opposed to only once in the amortized context of FreshDiskANN.
All these factors contribute to a significant lower performance for updates.

\begin{figure}
    \centering
    \includegraphics[width=0.45\textwidth]{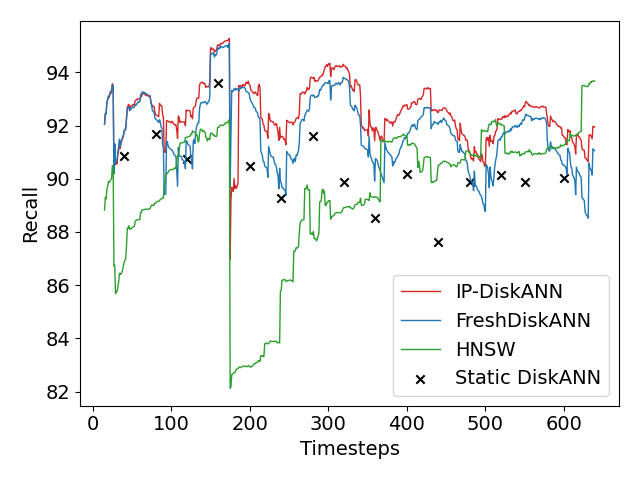}
    \vspace{-5pt}
    \caption{Recall comparison between FreshDiskANN, \name, HNSW, and ``rebuild the index from scratch'' approach (Static DiskANN) at every 64th step on the MSTuring-30M-clustered runbook.
    }
    \label{fig:build_scratch}
\end{figure}

To test the algorithm's scaling properties, we use the same type of runbook and dataset
and scale the number of datapoints from 1 to 10 to 35 million, yielding the runbooks
Wiki-1M-ExpirationTime, Wiki-10M-ExpirationTime, and Wiki-35M-ExpirationTime
(the last 3 rows in Figure~\ref{fig:highrecallplots} and Table~\ref{tab:running_time_highrecall}).
It is interesting to see that \name spends more time on deletion
for the 1 million scale dataset but becomes considerably faster as the scale of data
grows to 10 million and beyond. The reason is that most of 
the deletion time is spent searching for alternative edge connections,
the complexity of which grows sublinearly
in terms of the data size. On the other hand, FreshDiskANN scans through the
entire index trying to clean up each adjacency list, the complexity of which scales linearly.

We also compare the streaming approach with the approach of rebuilding the
index from scratch with the active point set at a given step.
This represents the default static approach to index construction
without a bound on computational resources.
For each selected time step,
we scan the runbook to identify all active points at the current step.
We then build a standard DiskANN index from scatch with the active points and measure 
the recall with the given parameters.
Due to limited resources, we report snapshots
from the MSTuring-30M-Clustered runbook every 64 steps in Figure~\ref{fig:build_scratch}.
It is interesting to observe that ``building from scratch'' does not yield
the optimal recall even compared to the FreshDiskANN baseline.
We hypothesize that deleting and adding new edges over time gradually
improves the graph connectivity which makes the graph easier to traverse than the initial graph.

We also report the performance the algorithms in a resource constrained environment,
where we accept lower recall for faster index construction.
Measurements in Figure~\ref{fig:lowrecallplots} and Table~\ref{tab:running_time_lowrecall}
show that even in this lower recall regime, \name outperforms the FreshDiskANN baseline, 
achieving an average recall improvement of 1.2, 1.8, and 5.2 across these runbooks while using fewer distance computations 
and providing higher query throughput. 
Also, our algorithm requires less time to handle deletions on the two larger runbooks. 
Although it takes more time on the Wiki-1m-ExpirationTime runbook, we believe this is due to the data size, as we have demonstrated that it scales well on larger datasets. (Cf. Table~\ref{tab:running_time_highrecall}).

\begin{table}[t]
\footnotesize
\begin{tabular}{p{2cm}|c|c|c|c|c}
\toprule
                                       &                 & Deletion & Insertion & Search & Recall@10 \\ 
\midrule
\multirow{2}{*}{\shortstack{MSTuring-10m \\ SlidingWindow}}           & FreshDiskANN         & 720     & 466      & 47    & 76.7   \\  
                                       & \name & 394     & 473      & 46    & 77.9   \\
\midrule
\multirow{2}{*}{\shortstack{MSTuring-30m \\ Clustered}} & FreshDiskANN         & 1376     & 1027      & 288    & 71.3   \\
                                       & \name & 1170     & 1031      & 288    & 73.1   \\

\midrule

\multirow{2}{*}{\shortstack{Wiki-1m \\ ExpirationTime}}                & FreshDiskANN         & 11      & 31       & 25    & 83.1      \\  
                                       & \name & 53       & 33       & 26     & 88.3   \\ 

\bottomrule
\end{tabular}
\caption{Comparison of running time between \name
and FreshDiskANN in the low-recall regime with $R=32, l_b=l_s=64$.
We report Recall@10, the total amount of time spent (in seconds)
on deletion (including consolidation), insertion, and executing queries across different runbooks.
\vspace{-10pt}
}
\label{tab:running_time_lowrecall}
\end{table}


\afterpage









%% file: highrecallplots.tex
\begin{figure*}[ht!]
    \centering
    \rotatebox{90}{\quad MSTuring-10M-SlidingWindow}
    \includegraphics[width=0.32\textwidth]{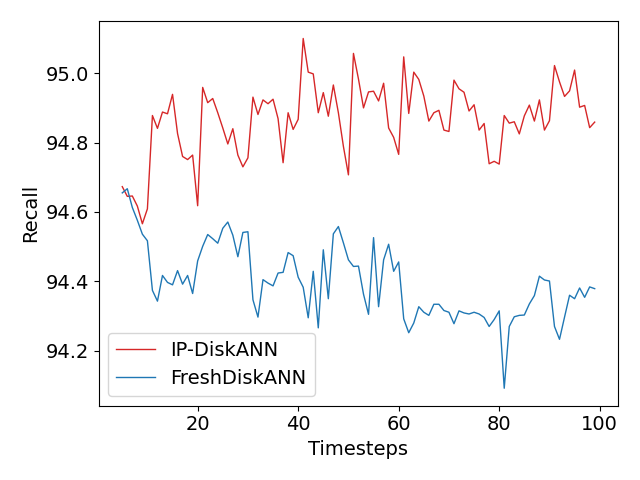}
    \hfill
    \includegraphics[width=0.32\textwidth]{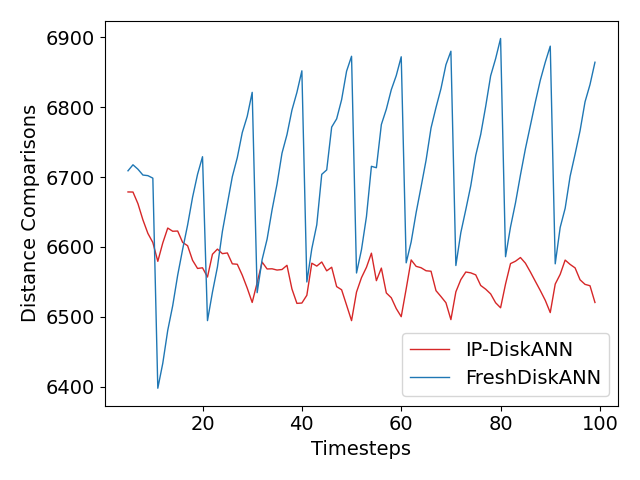}
    \hfill
    \includegraphics[width=0.33\textwidth]{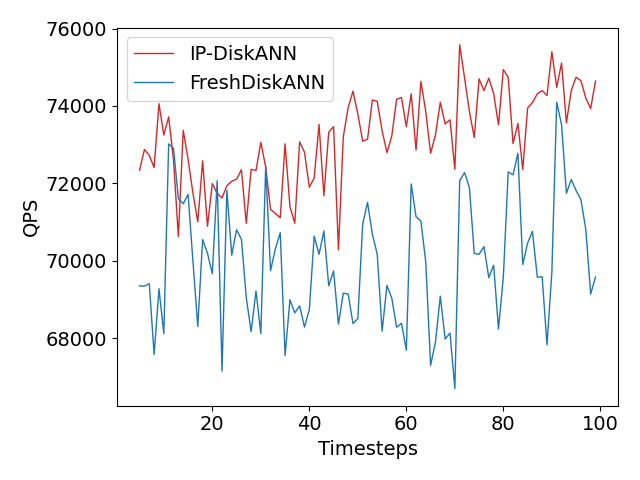}
    \vspace{-10pt}
    
    \centering
    \rotatebox{90}{\quad\quad MSTuring-30M-Clustered}
    \includegraphics[width=0.32\textwidth]{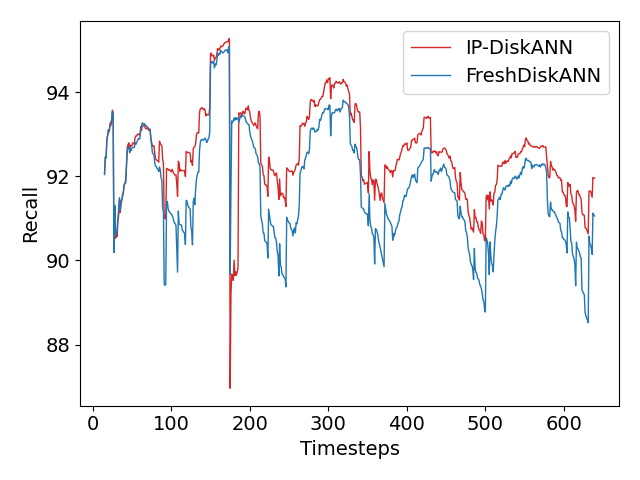}
    \hfill
    \includegraphics[width=0.32\textwidth]{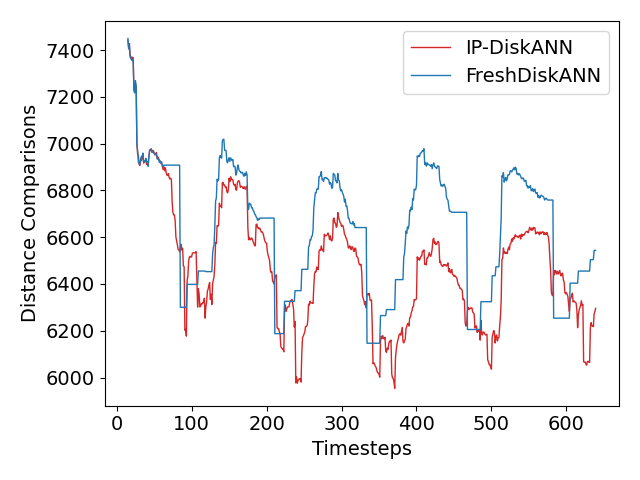}
    \hfill
    \includegraphics[width=0.32\textwidth]{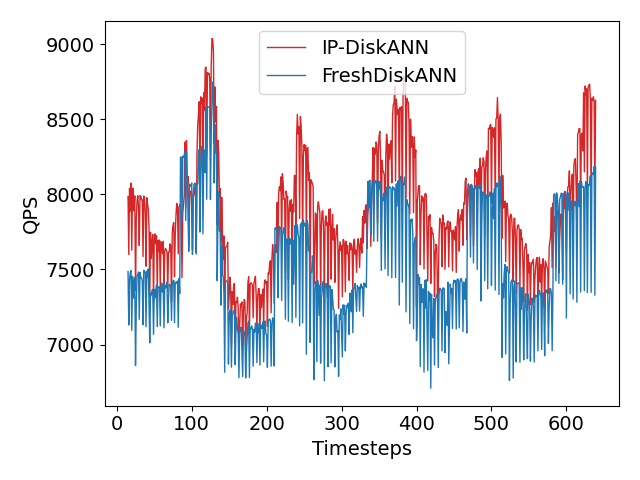}
    \vspace{-5pt}
    
    \centering
    \rotatebox{90}{\quad\quad Wiki-1M-ExpirationTime}
    \includegraphics[width=0.32\textwidth]{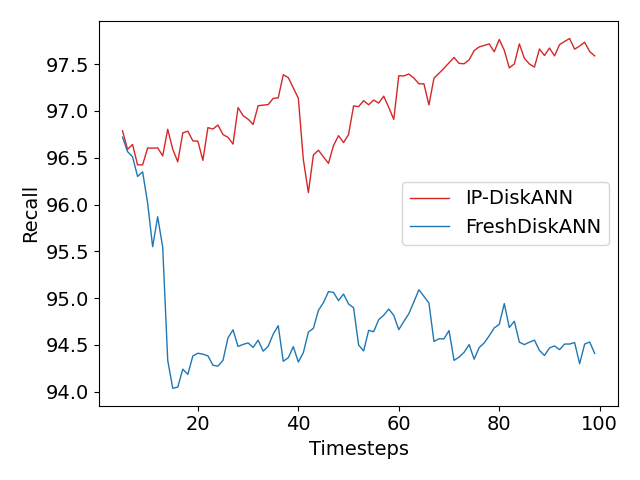}
    \hfill
    \includegraphics[width=0.32\textwidth]{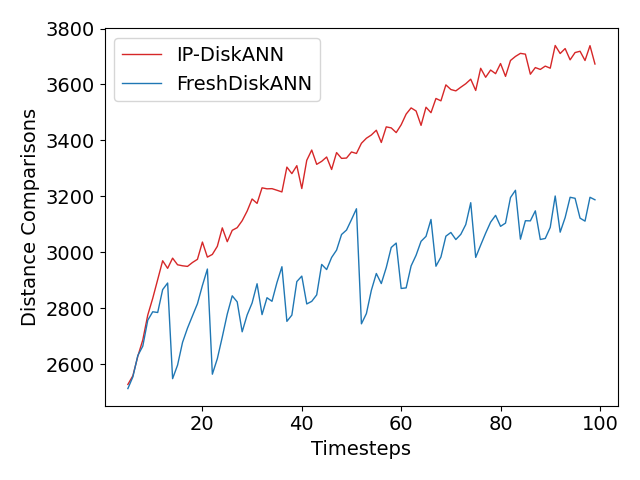}
    \hfill
    \includegraphics[width=0.32\textwidth]{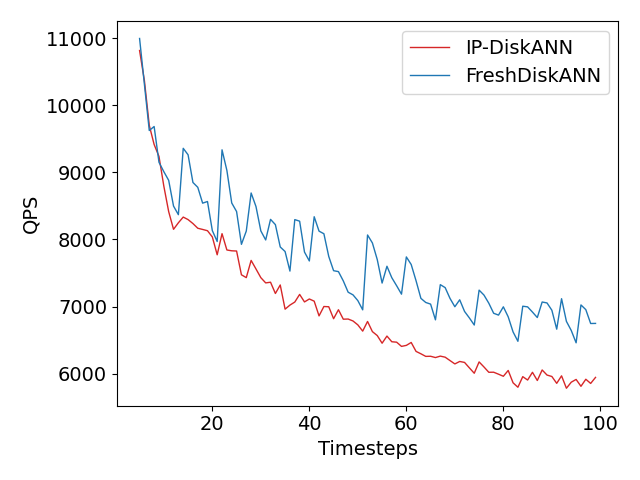}
    \vspace{-5pt}

    \centering
    \rotatebox{90}{\quad\quad Wiki-10M-ExpirationTime}
    \includegraphics[width=0.32\textwidth]{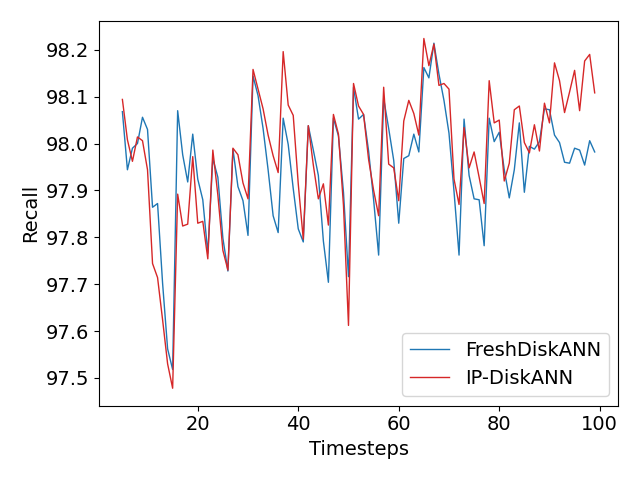}
    \hfill
    \includegraphics[width=0.32\textwidth]{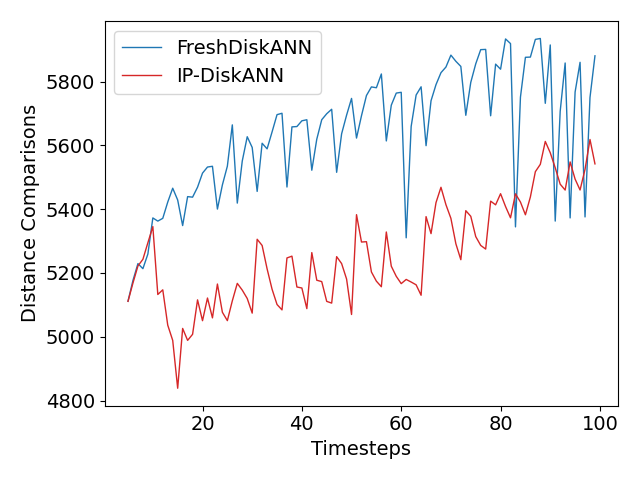}
    \hfill
    \includegraphics[width=0.32\textwidth]{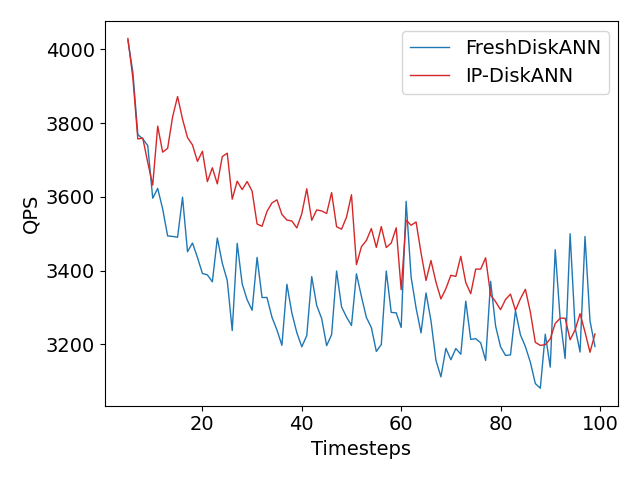}
    \vspace{-5pt}

    \centering
        \rotatebox{90}{\quad\quad Wiki-35M-ExpirationTime}
    \includegraphics[width=0.32\textwidth]{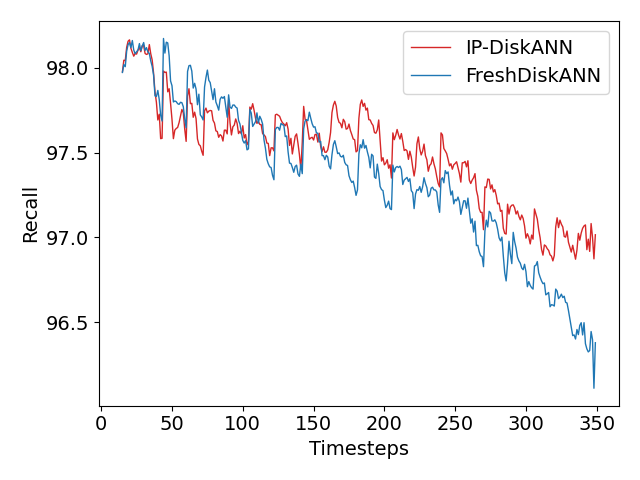}
    \hfill
    \includegraphics[width=0.32\textwidth]{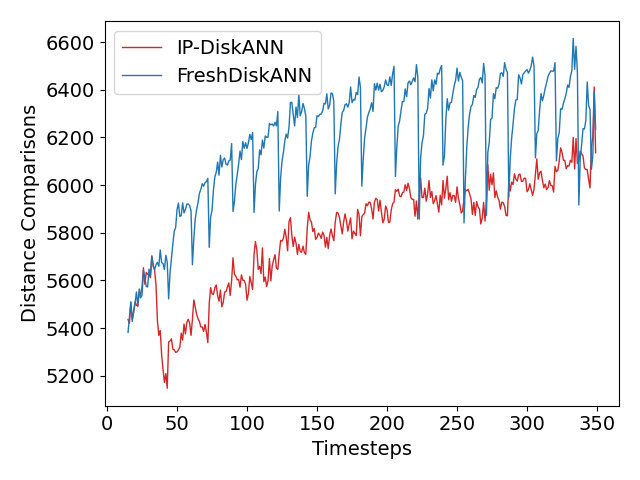}
    \hfill
    \includegraphics[width=0.32\textwidth]{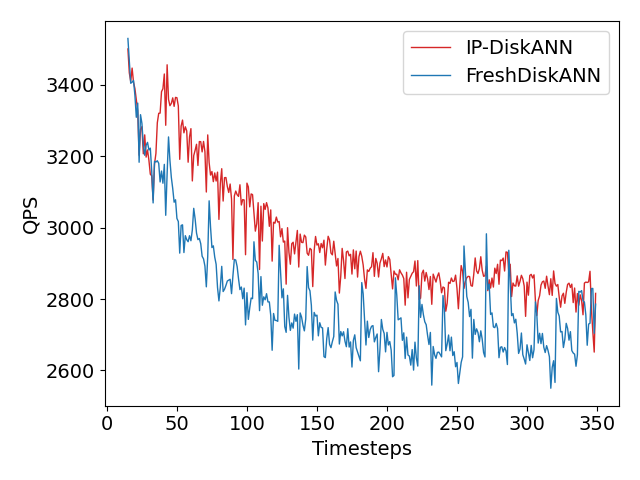}
    \vspace{-10pt}

    \caption{Comparison of recall (left), number of distance computations (middle), and queries per second (right)
    between \name and FreshDiskANN for various runbooks in the high-recall regime.
    with parameters $R=64, l_b=l_s=128$.
    }
    \label{fig:highrecallplots}
\end{figure*}

%% file: ablation.tex
\begin{figure*}[t]
    \centering
     \begin{subfigure}[t]{0.32\textwidth}
        \includegraphics[width=\linewidth]{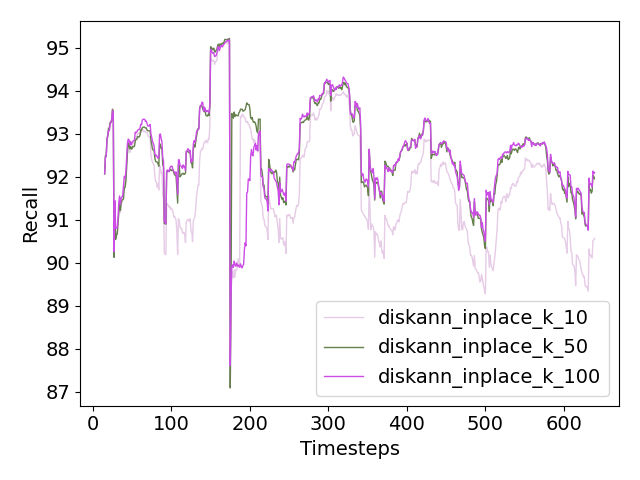}
        \caption{Influence of $k$.}
        \label{fig:ablation_k_plot}
    \end{subfigure}
~
     \begin{subfigure}[t]{0.32\textwidth}
        \includegraphics[width=\linewidth]{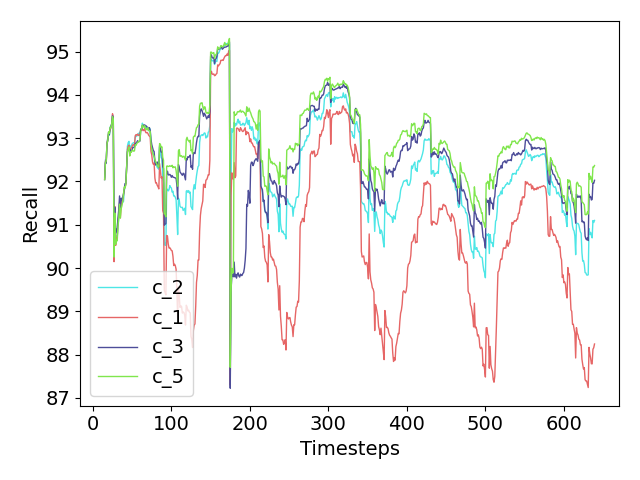}
        \caption{Influence of $c$.}
        \label{fig:ablation_c_plot}
    \end{subfigure}
~
     \begin{subfigure}[t]{0.32\textwidth}
        \includegraphics[width=\linewidth]{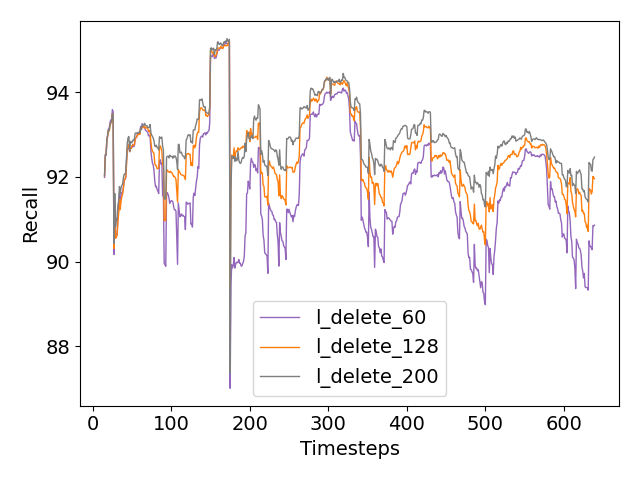}
        \caption{Influence of $l_{d}$.}
        \label{fig:ablation_k_plot}
    \end{subfigure}
        \vspace{-5pt}
    \caption{Recall@10 with in-place deletions at each step on MSTuring-30M-clustered
    with varying values of $k$, $c$ and $l_{d}$.}
\end{figure*}

\begin{table*}[h]
\centering
     \centering
     \begin{subtable}[t]{0.32\textwidth}
        \begin{tabular}{c|ccc}
            \toprule
              & k=10 & k=50 & k=100 \\
                \midrule
                Recall@10     & 91.8    & 92.5    & 92.6     \\
                Deletion time & 2983    & 3082    & 3099    \\
                \bottomrule
        \end{tabular}
        \caption{Influence of $k$.}
        \label{tab:ablation_k_table}
    \end{subtable}
~
    \begin{subtable}[t]{0.32\textwidth}
        \begin{tabular}{c|cccc}
            \toprule
                  & c=1 & c=2 & c=3 & c=5 \\
            \midrule
            Recall@10     & 91.0    & 92.2    & 92.5  & 92.8 \\
            Deletion time & 2400    & 2740    & 3148  & 3753 \\
            \bottomrule
        \end{tabular}
        \caption{Influence of $c$.}
        \label{tab:ablation_c_table}
     \end{subtable}
~~~
    \begin{subtable}[t]{0.32\textwidth}
        \begin{tabular}{c|ccc}
            \toprule
              & $l_d$=60 & $l_d$=128 & $l_d$=200 \\
            \midrule
            Recall@10     & 91.8    & 92.6    & 92.9   \\
            Deletion time & 1898    & 3160    & 4576   \\
            \bottomrule
        \end{tabular}
        \caption{Influence of  $l_d$.}
        \label{tab:ablation_l_delete_table}
    \end{subtable}
    \vspace{-5pt}
    \caption{Average recall (Recall@10) and total deletion time (in seconds)
        on the MSTuring-30M-clustered runbook.}
\end{table*}

\section{Ablation studies}
\label{sec:ablation}
In this section, we present ablation studies to examine the influence
of the parameters used in the in-place deletion algorithm. We vary the values of $k$,
$c$, $l_d$ in Algorithm~\ref{alg:deletion}, and the consolidation threshold $t$
in Algorithm~\ref{alg:consolidation_ours}.

As we increased the candidate list size $k$ from $10$ to $50$ to $100$, we found that the
average recall increases as expected (Figure~\ref{fig:ablation_k_plot}).
This is because retrieving more candidates makes it more likely the best
edges are found close to the deleted points and the estimated in-neighborhood
of a deleted point better aligns with the actual one. Additionally,
 more candidates to select replacement edges distributes the new edges across more nodes,
 reducing the likelihood of any single node triggering the pruning procedure. 
However, a larger $k$ requires more distance comparisons to select the top-$c$
closest points among the $k$ candidates, thereby increasing the deletion time (Table~\ref{tab:ablation_k_table}).

As we increase the number of replacement edges per delete by increasing
$c$ from $1$ to $5$, we found that the average recall increases significantly
from $c$=1 to $c$=2, but the benefit saturates as $c$ reaches $5$ (Figure~\ref{fig:ablation_c_plot}).
This  is expected, as introducing more edges increases graph connectivity.
The time spent on deletion increases almost linearly with $c$  (Table~\ref{tab:ablation_c_table}).
This is due to an increase in the number of prunes per delete, which in turn slows down the deletion process.
We found $c$=3 to be a good trade-off between accuracy and increased deletion time.

We varied the search parameter $l_d$ used inside in-place deletion between $60,128,200$.
We  found that the recall increases gradually with $l_d$ as the  larger beam size 
makes it more likely that the algorithm  finds the deleted point's in-neighbors.
However, a larger $l_d$ also significantly slows down the deletion function.
We observe that recall stops increasing when $l_d$ reaches $200$,
as the high maintenance cost outweighs the marginal increase in recall.

\begin{figure}
    \includegraphics[width=0.45\textwidth]{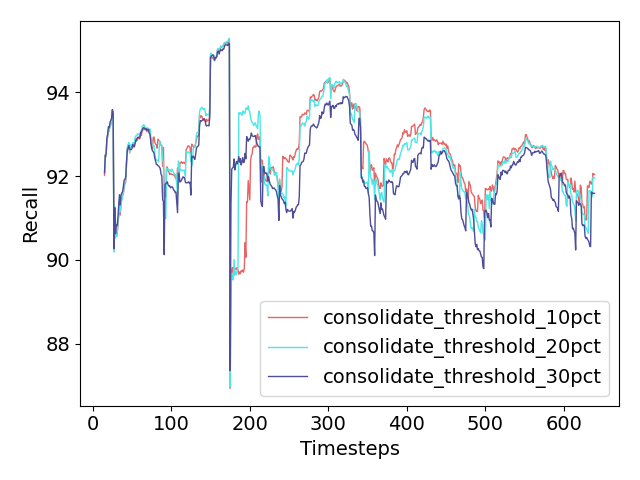}
    \vspace{-10pt}
    \caption{Recall@10 for varying consolidation threshold $t$ on the MSTuring-30M-Clustered runbook.}
    \label{fig:ablation_t_plot}
\end{figure}

\begin{table}[h]
\vspace{-10pt}
\begin{tabular}{c|ccc}
\toprule
 & t=30\% & t=20\% & t=10\% \\
\midrule
Recall        & 92.2    & 92.5    & 92.6  \\
Deletion time & 3136   & 3146    & 3207   \\
\bottomrule
\end{tabular}
\caption{Average recall (Recall@10) and total deletion time (in seconds) for various
values of consolidation threshold $t$ on the MSTuring-30M-clustered runbook.
\vspace{-10pt}}
\label{tab:ablation_t_table}
\end{table}

We varied the consolidation threshold to trigger our light-weight consolidation
in Algorithm~\ref{alg:consolidation_ours} between $t$=10\%/20\%/30\%.
The average recall increases as we perform consolidation more frequently (Figure~\ref{fig:ablation_t_plot}).
This is because fewer dangling edges make is less likely that the greedy search reaches dead ends.
At the same time, deletion time slightly increases as we consolidate more often (Table~\ref{tab:ablation_t_table}).


%% file: related.tex
\input{lowrecallplots}

\section{Related Work}
\label{sec:related}
The problem of K-Nearest Neighbor Search over a vector database
has been a popular research topic for many decades. 
Solutions are usually categorized by the data structure that stores points:
inverted file, search tree, hash table, or graph.
Since there are several recent surveys of past work on algorithms
and systems~\cite{AumullerCeccarello23, GraphANNSSurvey21, PanWL24, TAIPALUS2024101216, TianYZ0Z023},
we summarize the approaches briefly and concentrate on how they relate to \name,
which is the main subject of this paper.

In the inverted file approach, points are partitioned into 
clusters, each represented by its centroid.
A search for the nearest neighbors of a vector finds the centroids closest to the query vector and then scans the centroids' clusters to find the k-nearest vectors.

There are several strategies for processing updates: 
periodic rebuilds~\cite{AnalyticDBV}, which cause latency spikes; 
adding/deleting vectors from partitions with~\cite{ArandjelovicZ13} or without~\cite{pgvector} updating centroids, which can lead to unbalanced partitions and hence highly variable search times; 
compensating for drift by reclustering the largest and smallest partition (DeDrift~\cite{BaranchukDUY23});
splitting or merging clusters that exceed defined thresholds (SPFresh~\cite{YumingXuEtAl23});
and using the read intensity and partition imbalance to decide which partitions
to re-index and how often (Ada-IVF~\cite{MahoneyEtAl24}).
SPFresh  reports better performance than FreshDiskANN (Fig. 7 of~\cite{YumingXuEtAl23})
on a random runbook
that loosely corresponds to the ExpirationTime or SlidingWindow runbook
in this paper.
However, our experiments as well as  public benchmarks~\cite{bigann23}
demonstrate much higher and stable recall for FreshDiskANN on these runbooks
and for the more challenging Clustered runbook. 
This is possibly because experiments in SPFresh use an old fork [\href{https://github.com/Yuming-Xu/DiskANN_Baseline/commits/main/}{\textcolor{blue}{link}}] of DiskANN, 
rather than the official public release (v0.5.0.rc3
from \href{https://github.com/microsoft/DiskANN/tags}{\textcolor{blue}{link}}).
The experiments by Ada-IVF show it performs better than SPFresh on the clustered runbook,
but they do not compare to other algorithms in the NeurIPS'23 BigANN track~\cite{simhadri2024resultsbigannneurips23}. 
We leave an apples-to-apples comparison of these algorithms
based on normalized workloads to \name for future work.

Search trees are a partitioning method, where internal nodes
represent partitions and direct a query to nearby
partitions in logarithmic time.
K-d trees~\citeN{Bentley75}, Cover-tree~\cite{Beygelzimer06} and R*-trees~\cite{Beckmann90} 
are few examples, and updates to these indices have been  studied~\cite{dobson2021parallelnearestneighborslow}.
However, these methods are best suited to low dimensional spaces
and do not scale to hundreds of dimensions.

Locality-Sensitive hashing (LSH) relies on nearby points hashing to the same bucket~\cite{Indyk98}.
Each point is hashed into many hash tables using different functions.
Numerous variations have been devised, see~\cite{LSHSurvey08}.
However, they require a lot of memory and perform worse than the other methods on disk-resident data.
Parallel LSH~\cite{PLSH13} improves LSH performance via innovations in
concurrent and cache-friendly hash construction and search. 
Streaming updates are handled via buffering in a delta table
which is merged periodically into the main structure.
This extension mirrors incremental merges in proximity graphs~\cite{freshdiskann}.
However, graph-based methods achieve the scale and query throughput with a lot fewer machines than PLSH.


All proximity graph algorithms use a greedy search algorithm, such as Algorithm~\ref{alg:search}.
They primarily differ in their method for graph construction.
In addition to DiskANN's construction algorithm, HNSW~\cite{HNSW16}, NSG~\cite{NSG17}, and HCNNG~\cite{MunozGDT19} have been shown to perform well. 
In this paper, we focused on DiskANN because it has won benchmark competitions,
performs well for both in-memory and on-disk indices, and is used in commercial systems.
It is possible our algorithm lends itself to be incorporated into other graph indices---a topic for future work.

The only in-place delete algorithm for DiskANN that we know of is described in~\cite{ZhaozhuXuEtAl22}.
Unlike our solution, it requires that each vertex has pointers to its in-neighbors, which is expensive as noted earlier.
For each in-neighbor of the deleted vertex $x_j$, it replaces $x_j$'s out-edges by pointers to a diverse set of its nearest neighbors, that is, neighbors that are closer to $x_j$ than they are to each other.
Their experiments show that for a given recall goal, their algorithm produces a graph that executes queries faster than other heuristics and often faster than a fresh rebuild of the graph.

Sinnamon~\cite{BruchNIL24} is an algorithm for streaming ANNS over sparse vectors with thousands to millions of dimensions, i.e., where very few dimensions of each vector are non-zero.
None of the above data structures work well for this case, so a specialized one is required.
RTAMS-GANNS~\cite{YipingSunEtAl24} implements ANNS on GPUs.
But unlike our work, it only handles insertions, not deletions.

%% file: lowrecallplots.tex
\begin{figure*}[t]
    \centering
    \rotatebox{90}{\quad MSTuring-10m-slidingwindow}
    \includegraphics[width=0.32\textwidth]{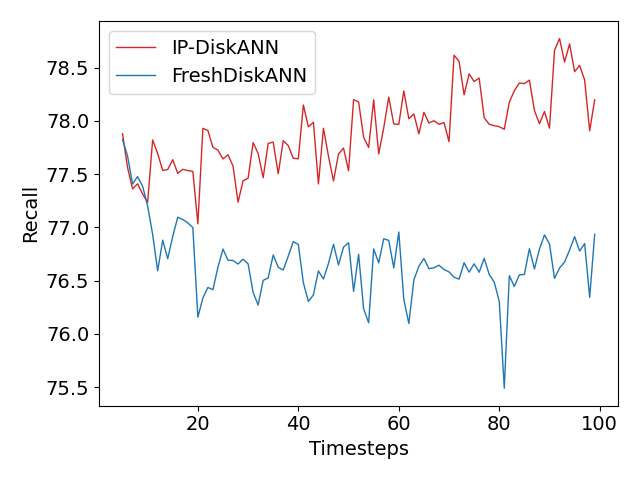}
    \hfill
    \includegraphics[width=0.32\textwidth]{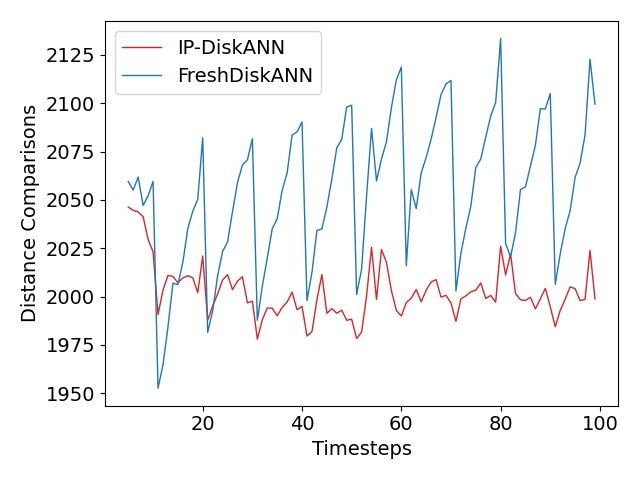}
    \hfill
    \includegraphics[width=0.32\textwidth]{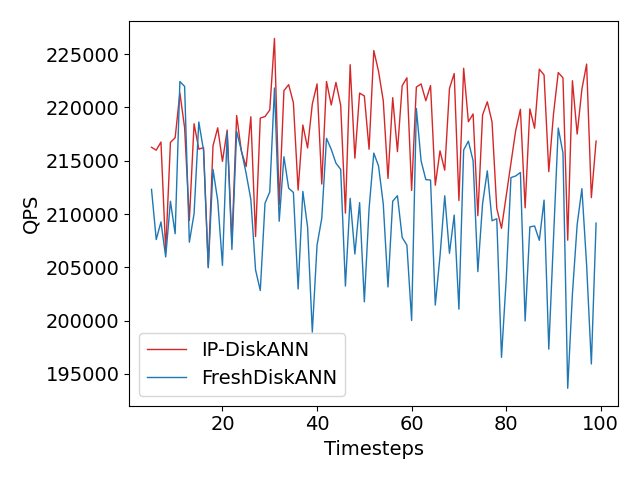}

    \centering
    \rotatebox{90}{\quad MSTuring-30m-clustered}
    \includegraphics[width=0.32\textwidth]{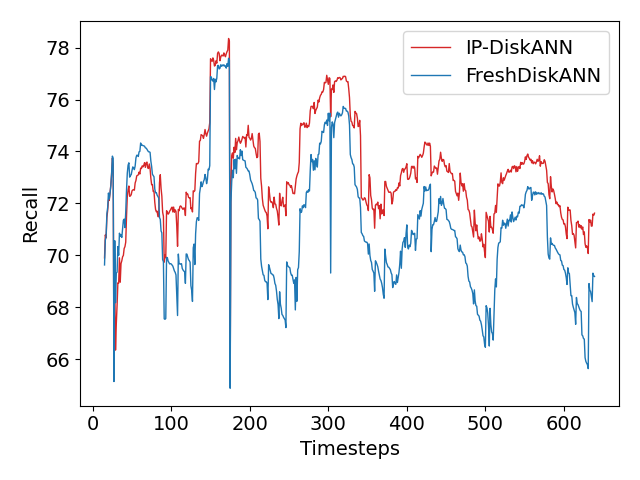}
    \hfill
    \includegraphics[width=0.32\textwidth]{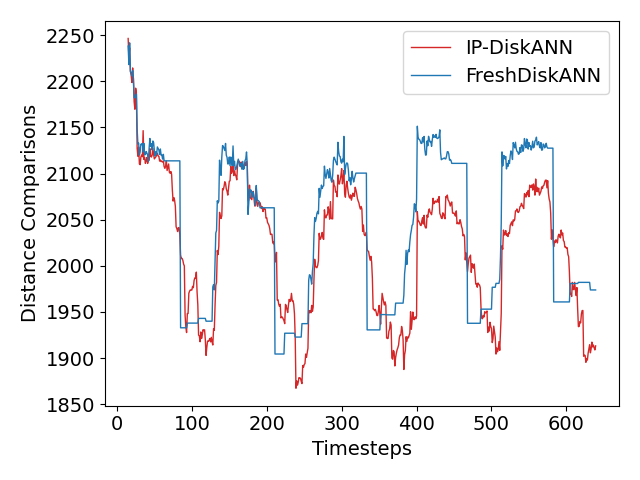}
    \hfill
    \includegraphics[width=0.32\textwidth]{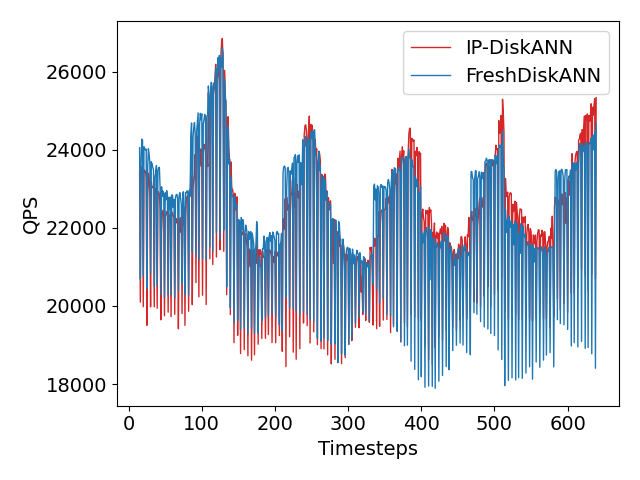}

    \centering
    \rotatebox{90}{\quad Wiki-1M-ExpirationTime}
    \includegraphics[width=0.32\textwidth]{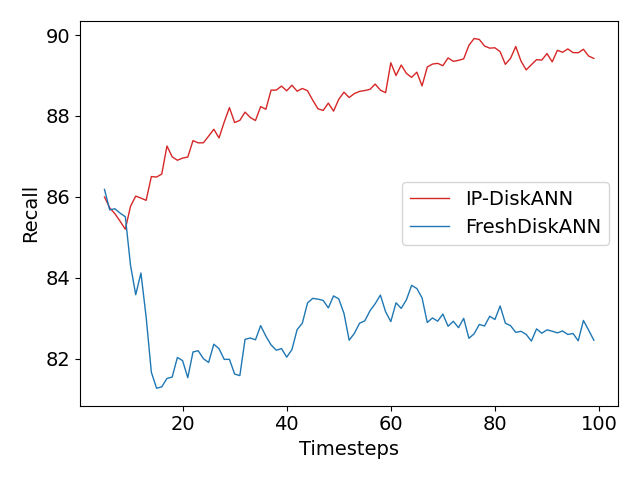}
    \hfill
    \includegraphics[width=0.32\textwidth]{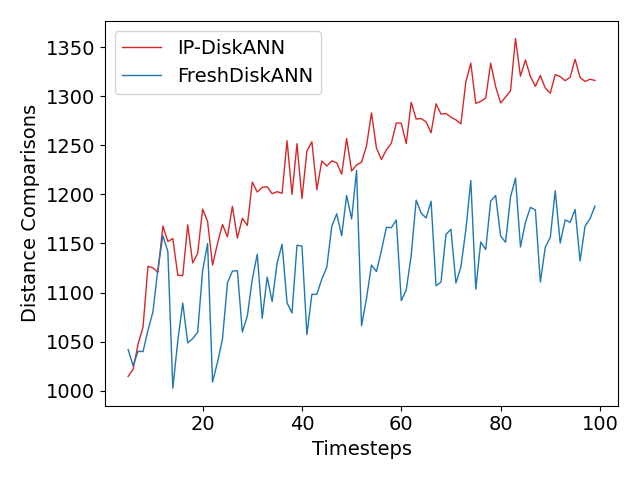}
    \hfill
    \includegraphics[width=0.32\textwidth]{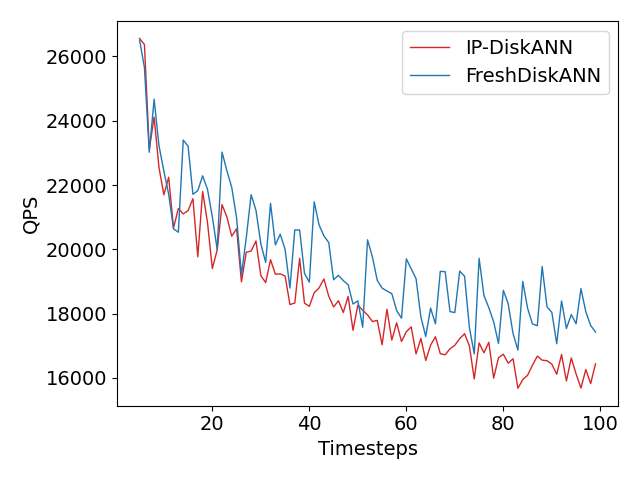}
    \vspace{-7pt}
    \caption{Comparison of recall (left), number of distance computations (middle),
    and queries per second (right)
    between \name  and FreshDiskANN for 
    various runbooks in the low-recall regime using parameters $R=32, l_b=l_s=64$.}
    \label{fig:lowrecallplots}
\end{figure*}

%% file: conclusion.tex
\section{Conclusion}
\label{sec:conclusion}

In this paper, we introduced \name, the first in-place deletion algorithm for DiskANN
over the common case of a singly-linked proximity graph. 
\name has stable recall over a variety of update patterns and better
query and update complexity compared to previous graph-based indices.
It avoids usage spikes and query inefficiencies of past approaches that
collect updates in delta files that are periodically consolidated into the primary index.

Future work includes finding other adversarial workloads to test robustness,
such as one for replacements, and an experimental comparison of \name with the
best streaming methods for other data structures, such as IVF and LSH.